\documentclass[aps,prd,onecolumn,eqsecnum,amsmath,nofootinbib,preprintnumbers]{revtex4}%
\newcounter{example}[section]

\usepackage{graphicx}
\usepackage{hyperref}
\usepackage{ulem}
\usepackage{amsmath}
\usepackage{color,graphicx,float,subfigure}%[dvips]%
\usepackage{amsfonts,amssymb,theorem,mathrsfs,times}
\usepackage{bm}
\usepackage{amsmath}
{\theorembodyfont{\upshape}
	}
{\theorembodyfont{\upshape}
	}
{\theorembodyfont{\upshape}
	}
{\theorembodyfont{\upshape}
	}
{\theorembodyfont{\upshape}
	}
{\theorembodyfont{\upshape}
	}

\newcommand{\dalm}{\kern1pt\vbox{\hrule height 0.9pt\hbox{\vrule width
			0.9pt\hskip 2.5pt\vbox{\vskip 5.5pt}\hskip 3pt\vrule width
			0.3pt}\hrule height 0.3pt}\kern1pt}

\begin{document}
	
	\title{Static spheres in black hole spacetimes: pairing, energy conditions, and an upper bound on the innermost radius}
	
	\author{Yong Song\footnote{e-mail address: syong@cdut.edu.cn}}
	\affiliation{College of Physics, Chengdu University of Technology, Chengdu, Sichuan 610059, China}
	
	\date{\today}
	
	\begin{abstract}
		Static spheres---spherical surfaces, located outside the event horizon, on which massive test particles can remain at rest with zero angular momentum relative to asymptotic static observers---have recently attracted some attention. In this work, we investigate their existence, relation to the energy conditions, and radial bounds in general static, spherically symmetric, asymptotically flat black hole spacetimes within general relativity. By analyzing the behavior of a radial function constructed from the mass and radial pressure functions, we prove that, under the weak energy condition (WEC), a static sphere necessarily requires negative radial pressure (tension). Furthermore, for non-extremal black holes satisfying the WEC, we show that any exterior static spheres, if present and non-degenerate, occur in pairs, each consisting of an inner unstable sphere and an outer stable one. Assuming the WEC holds throughout the exterior, the inner static sphere is characterized by a violation of the strong energy condition (SEC), with $\rho+p+2p_T<0$, where $\rho$, $p$, and $p_T$ denote the energy density, radial pressure, and tangential pressure, respectively; the SEC is restored ($\rho+p+2p_T>0$) at the outer sphere; a degenerate static sphere satisfies $\rho+p+2p_T=0$. In addition, focusing on the innermost static sphere and assuming that the WEC holds while the SEC is uniformly violated between the event horizon and this sphere, we derive a rigorous upper bound on its radius. These results establish a direct, analytic link between the energy conditions and the existence of static spheres, and provide a quantitative constraint on the matter environment of black holes possessing such orbits.
	\end{abstract}
	
	\maketitle
	
	\section{Introduction}
	
	The study of particle orbits in black hole spacetimes serves as a fundamental probe of the geometry and matter content of the strong-field regime. Among the various types of orbits, static spheres---spherical surfaces on which massive test particles can remain at rest with zero angular momentum relative to asymptotic static observers---have recently attracted some attention. More precisely, in a static spacetime, a static sphere can be characterized by the existence of free-falling massive particles whose four-momentum is parallel to the asymptotically timelike Killing vector field. In a vacuum Schwarzschild spacetime, such static equilibrium is impossible: a particle with vanishing angular momentum inevitably falls into the horizon or escapes to infinity. However, in certain rotating non-Kerr black holes, a static point can exist in the equatorial plane through appropriate tuning of the angular momentum of counter-rotating particles~\cite{Collodel:2017end}; by axial symmetry, this extends to a ring of static points. The existence of these or related static configurations therefore provides a potential observational signature of compact objects that deviate from the Kerr metric~\cite{Collodel:2021gxu,Teodoro:2021ezj,Zhang:2021xhp}.
	
	Static spheres can indeed exist when non-trivial matter fields or modifications to gravity are present~\cite{Liu:2019rib,Wei:2023bgp,Yerra:2024stj,Zhao:2026cne}. In Ref.~\cite{Wei:2023bgp}, Wei et al. employed a topological approach to study the properties of static spheres in spherically symmetric spacetimes. They showed that when gravity is coupled to quasi-topological electromagnetism, the resulting effective repulsive force can balance the gravitational attraction, thereby allowing for static equilibrium points. Due to spherical symmetry, these points form entire static spheres. Since the outer static spheres are stable---a test particle slightly displaced from one oscillates about the equilibrium rather than falling in or escaping---they provide, in principle, a natural supporting locus for thin shells of matter (or for artificial megastructures) encircling a black hole, in analogy with a Dyson sphere~\cite{Dyson:1960xib}. More generally, a static sphere is a precise local diagnostic of the spacetime's matter content: it is a surface on which a free test particle at rest feels no net radial acceleration, a situation that is impossible in vacuum Schwarzschild and can only arise when non-trivial matter (or modified gravity) is present. Its existence and location therefore encode properties of the energy--momentum tensor in the strong-field region, making static spheres valuable both as a probe of non-trivial matter fields and as a potential observational target. Furthermore, their topological analysis provided a powerful global argument: for an asymptotically flat black hole, the total topological number $W$ associated with static spheres is zero, implying that static spheres appear in pairs with opposite stability indices (one stable and one unstable). In contrast, naked singularities belong to a different topological class ($W=1$) and can host an odd number of static spheres, as exemplified in the Reissner-Nordstr\"{o}m spacetime~\cite{Pugliese:2010ps}. Despite these insights, fundamental questions remain largely open: Under what local conditions do static spheres exist outside black hole horizons? And what does the existence of a static sphere directly imply for the matter field in its vicinity?
	
	On the other hand, universal bounds on circular orbits have attracted considerable attention in recent years. For null circular geodesics---photon spheres---it has been proved that, under suitable energy conditions, every static, spherically symmetric, asymptotically flat black hole spacetime admits a photon sphere whose radius satisfies $r_\gamma\le 3M$, where $M$ is the ADM mass, with the Schwarzschild black hole saturating this bound~\cite{Hod:2013jhd,Yang:2019zcn}. This result has since been generalized to higher dimensions~\cite{Song:2026cci} and extended to stable photon spheres~\cite{Song:2025fgq}. Complementing these spherically symmetric results, important existence and stability theorems for light rings---including the presence of unstable and stable fundamental photon orbits---have been established for stationary, axisymmetric spacetimes that need not be spherically symmetric~\cite{Cunha:2017wwe,Cunha:2020ktt}. Universal bounds have also been investigated for compact stars under different energy conditions~\cite{Peng:2018nkj,Liu:2024odv}, as well as for photon spheres in modified gravity theories~\cite{Lu:2019zxb,Cvetic:2016bxi,Ma:2019ybz}. Besides these upper bounds, lower bounds on the radii of photon spheres have also been established under additional physically motivated assumptions~\cite{Yang:2019zcn,Hod:2020pim,Hod:2023jmx,Song:2026cci}. For timelike circular orbits, particular attention has been devoted to the innermost stable circular orbit (ISCO), which marks the inner edge of accretion disks. A universal upper bound $r_{\mathrm{ISCO}}\le 6M$ has recently been derived for static, spherically symmetric, asymptotically flat black holes under appropriate energy conditions~\cite{Cen:2025atm}, with Schwarzschild again saturating the inequality. 
	
	In contrast to the extensive literature on photon spheres and ISCOs, a systematic understanding of the radial bounds of static spheres---specifically, whether their locations are subject to universal upper bounds analogous to those for photon spheres and ISCOs---has remained elusive. The present work aims to fill this gap.
	
	In this work, we first address the local conditions for static spheres by analyzing the spacetime geometry and the energy-momentum tensor, without resorting to global topological arguments. We introduce a radial function $\mathcal{N}(r)$ constructed from the mass and radial pressure functions, and analyze its behavior for $r\ge r_H$, where $r_H$ is the horizon radius. Our analysis reveals that, under the WEC, a static sphere necessarily requires a negative radial pressure (tension). Moreover, for a non-extremal black hole satisfying the WEC, we prove that static spheres appear in pairs with opposite radial stability (one unstable inner, one stable outer), provided that the static spheres outside the horizon are finite in number and all are non-degenerate. We uncover the underlying physical origin: in the paired case, assuming the Weak Energy Condition (WEC) always holds, the inner (unstable) static sphere is characterized by a violation of the SEC, with $\rho+p+2p_T<0$, while at the outer (stable) one the SEC is restored ($\rho+p+2p_T>0$); a degenerate static sphere satisfies $\rho+p+2p_T=0$. These results establish a direct, analytic connection between the energy conditions and the existence of static spheres, providing a novel local diagnostic tool for probing the matter content in the strong-field regions of black hole spacetimes.
	
	Then, focusing on the innermost static sphere and assuming that the weak energy condition holds while the strong energy condition is uniformly violated between the event horizon and the static sphere, we derive a rigorous upper bound on its radius,
	\[
	r^-_{\mathrm{sp}}\le \left[r_H^3+\frac{3r_H(1-8\pi r^2_H\rho(r_H))}{8\pi\kappa}\right]^{1/3},
	\]
	where $\rho(r_H)$ is the energy density at the horizon, and $\kappa$ is a positive uniform lower bound on $-(\rho+p+2p_T)$ between $r_H$ and $r^-_{\mathrm{sp}}$. This inequality quantifies how the horizon deficit and the strength of the SEC violation jointly limit the region where a static sphere can exist.
	
	The paper is organized as follows: In Sec.~\ref{section2}, we set up the general metric for a static, spherically symmetric black hole and derive the Einstein field equations and the conservation equation for the energy-momentum tensor. In Sec.~\ref{section3}, we derive the condition for the existence of static spheres, analyze the behavior of $\mathcal{N}(r)$, and prove the main results regarding the relationship between static spheres and the SEC. In Sec.~\ref{section4}, we introduce an auxiliary function and derive the upper bound on the radius of the innermost static sphere. Finally, Sec.~\ref{conclusion} contains a summary and discussion of the implications of our results.
	
	\section{Description of the Black Hole Spacetime}\label{section2}
	
	The line element of a static, spherically symmetric, and asymptotically flat black hole spacetime can be expressed as~\cite{Hod:2013jhd,Hod:2020pim,Song:2025fgq,Cen:2025atm}
	\begin{align}
		\label{dugui}
		ds^2=-e^{-2\delta}\mu dt^2+\mu^{-1}dr^2+r^2(d\theta^2+\sin^2\theta d\phi^2)\;,
	\end{align}
	where the metric functions $\mu(r)$ and $\delta(r)$ depend only on the radial coordinate $r$. Asymptotic flatness demands that, in the limit $r\rightarrow \infty$,
	\begin{align}
		\label{infinity}
		\mu(r\rightarrow \infty)\rightarrow1\quad\mathrm{and}\quad\delta(r\rightarrow \infty)\rightarrow 0\;.
	\end{align}
	We do not impose $\delta(r)=0$, a condition equivalent to $g_{tt}g_{rr}=-1$~\cite{Jacobson:2007tj}, so our analysis remains valid for hairy black hole configurations as well~\cite{Volkov:1998cc,Volkov:2016ehx}.
	
	We denote the components of the energy-momentum tensor as $T^t_t=-\rho$, $T^r_r=p$ and $T^{\theta}_{\theta}=T^{\phi}_{\phi}=p_T$, where $\rho$, $p$ and $p_T$ are the energy density, radial pressure, and tangential pressure, respectively. The Einstein field equations $G^{\mu}_{\nu}=8\pi T^{\mu}_{\nu}$ yield
	\begin{align}
		\label{mu1}
		&\mu'=-8\pi r \rho+\frac{1-\mu}{r}\;,\\
		\label{delta1}
		&\delta'=-\frac{4\pi r(\rho+p)}{\mu}\;,
	\end{align}
	where a prime denotes differentiation with respect to $r$. 
	
	Regularity at the event horizon $r=r_H$ requires
	\begin{align}
		\label{murH}
		\mu(r_H)=0\quad\mathrm{with}\quad\mu'(r_H)\ge 0\;,
	\end{align}
	and
	\begin{align}
		\label{deltarH}
		\delta(r_H)<\infty\quad ;\quad \delta'(r_H)<\infty\;.
	\end{align}
	From Eqs.~(\ref{mu1}) and (\ref{murH}) we obtain the boundary condition
	\begin{align}
		\label{rhorH}
		8\pi r^2_{H}\rho(r_H)\le 1\;,
	\end{align}
	where the equality corresponds to an extremal black hole. Eqs.~(\ref{delta1}) and (\ref{deltarH}) give the additional condition
	\begin{align}
		\label{prH}
		-p(r_H)=\rho(r_H)\;.
	\end{align}
	
	The mass $m(r)$ enclosed within a sphere of radius $r$ is defined by
	\begin{align}
		\label{m}
		m(r)=\frac{1}{2}r_H+\int_{r_H}^{r}4\pi r'^2\rho(r')dr'\;,
	\end{align}
	where $r_H$ is the radius of the event horizon, and the horizon mass $m(r_H)=r_H/2$.  The ADM mass of the spacetime is $M \equiv m(r\to\infty)$.
	
	Combining Eqs.~(\ref{mu1}) and (\ref{m}) yields the relation between $\mu$ and the mass $m(r)$,
	\begin{align}
		\label{mu=m}
		\mu(r)=1-\frac{2m(r)}{r}\;.
	\end{align}
	From Eq.~(\ref{mu1}), we also have
	\begin{align}
		\label{m1}
		m'(r)=4\pi r^2\rho\;.
	\end{align}
	In addition to a finite ADM mass, we assume the asymptotic falloff
	\begin{align}
		\label{r3rho}
		r^3\rho(r)\to 0\quad \mathrm{as} \quad r\to\infty\;.
	\end{align}
	We also assume that
	\begin{align}
		\label{r3p}
		r^3p(r)\to 0\quad \mathrm{as} \quad r\to\infty\;.
	\end{align}
	
	The conservation equation for the energy-momentum tensor has only one non-trivial component,
	\begin{align}
		\label{Tmunu}
		T^{\mu}_{r;\mu}=0\;.
	\end{align}
	Substituting Eqs.~(\ref{mu1}) and (\ref{delta1}) into Eq.~(\ref{Tmunu}) gives the radial pressure gradient
	\begin{align}
		\label{p1}
		p'(r)=\frac{-p-8\pi r^2 p^2-5p\mu+2T\mu-\rho-8\pi r^2 p\rho+3\mu \rho}{2r\mu}\;,
	\end{align}
	where 
	\begin{align}
		\label{T}
		T=-\rho +p+2p_T
	\end{align}
	is the trace of the energy-momentum tensor. In the subsequent analysis we will invoke the Weak Energy Condition (WEC) and the Strong Energy Condition (SEC). The WEC and the SEC are given by~\cite{Guo:2022ghl}:
	\begin{itemize}
		\item[(1)] Weak Energy Condition (WEC): The energy density is positive semidefinite,
		\begin{align}
			\label{wec}
			\rho\ge 0\;,
		\end{align}
		and it bounds the pressures, which implies the inequalities
		\begin{align}
			\label{wec1}
			\rho+p\ge 0 \quad \text{and} \quad \rho+p_T\ge 0 \;.
		\end{align}
		
		\item[(2)] Strong Energy Condition (SEC): The SEC requires the matter fields satisfy the following inequalities
		\begin{align}
			\label{sec}
			\rho + p +2p_T \ge 0 \quad \text{and} \quad \rho+p\ge 0,\quad \rho+p_T\ge 0\;.
		\end{align}
	\end{itemize}
	
	We note that, under the WEC, the two inequalities $\rho+p\ge0$ and $\rho+p_T\ge0$ entering Eq.~(\ref{sec}) are already guaranteed by Eq.~(\ref{wec1}); the SEC therefore reduces to the single condition $\rho+p+2p_T\ge0$. Accordingly, throughout this work ``SEC violation'' will be understood as $\rho+p+2p_T<0$. Matter satisfying the WEC but violating the SEC is by no means exotic: a positive cosmological constant, with $\rho=\Lambda/(8\pi)>0$ and $p=p_T=-\Lambda/(8\pi)$, gives $\rho+p+2p_T=-\Lambda/(4\pi)<0$ (although a cosmological constant extending to infinity is not asymptotically flat); likewise, a slowly evolving massive scalar field, whose energy density is dominated by its (positive) potential, has $\rho\simeq V$, $p\simeq -V$, and hence $\rho+3p\simeq -2V<0$.  Such sources arise routinely in models of dark energy~\cite{Copeland:2006wr}, inflation~\cite{Linde:2000kn}, and scalar-tensor theories~\cite{Fujii:2003pa}.
	
	\section{Static Spheres and Strong Energy Condition Violation}\label{section3}
	
	Due to the spherical symmetry of the system, we restrict the analysis to the equatorial plane $\theta=\frac{\pi}{2}$. The Lagrangian governing geodesics in the spacetime (\ref{dugui}) is
	\begin{align}
		\label{la}
		2\mathcal{L}=-e^{-2\delta}\mu \dot{t}^2+\mu^{-1}\dot{r}^2+r^2\dot{\phi}^2=\epsilon \;,
	\end{align}
	where $\epsilon=-1,0,+1$ for timelike, null and spacelike geodesics, respectively. For static spheres, we consider timelike geodesics ($\epsilon=-1$). Because the metric (\ref{dugui}) is independent of $t$ and $\phi$, the specific energy $E$ and specific angular momentum $L$ are conserved along the geodesic.
	
	The generalized momenta derived from (\ref{la}) are
	\begin{align}
		\label{nengliang}
		&p_t=-e^{-2\delta}\mu\dot{t}=-E\;,\\
		\label{jiaodongliang}
		&p_{\phi}=r^2\dot{\phi}=L\;.
	\end{align}
	Inserting Eqs.~(\ref{nengliang}) and (\ref{jiaodongliang}) into Eq.~(\ref{la}) leads to
	\begin{align}
		\dot{r}^2+V_{\mathrm{eff}}=0\;,
	\end{align}
	where the effective potential for timelike geodesics is
	\begin{align}
		V_{\mathrm{eff}}=-\frac{E^2}{e^{-2\delta }}+\frac{L^2\mu}{r^2}+\mu\;.
	\end{align}
	A static sphere is a circular orbit with zero angular momentum ($L=0$), satisfying $\dot{r}=0$ and $\ddot{r}=0$~\cite{Wei:2023bgp}. Setting $L=0$, the effective potential reduces to
	\begin{align}
		\mathcal{V}_{\mathrm{eff}}=-\frac{E^2}{e^{-2\delta }}+\mu\;.
	\end{align}
	The circular orbit conditions $\mathcal{V}_{\mathrm{eff}}=0$ and $\mathcal{V}_{\mathrm{eff}}'(r)=0$ lead to
	\begin{align}
		\label{N=0}
		\mathcal{N}(r=r_{\mathrm{sp}})=0\;,
	\end{align}
	where $r_{\mathrm{sp}}>r_H$ , and
	\begin{align}
		\label{N}
		\mathcal{N}(r)=\mu(r)-1-8\pi r^2 p(r)=-\frac{2}{r}[m(r)+4\pi r^3 p(r)]\;,
	\end{align}
	with the second equality following from (\ref{mu=m}). One can also obtain the relation
	% --- Corrected equation (sign changed) ---
	\begin{align}
		\label{VNrelation}
		\mathcal{V}''_{\mathrm{eff}}(r_{\mathrm{sp}})=-\frac{1}{r_{\mathrm{sp}}}\mathcal{N}'(r_{\mathrm{sp}})\;,
	\end{align}
	which determines the linear radial stability of a non-degenerate static sphere. $\mathcal{V}''_{\mathrm{eff}}(r_{\mathrm{sp}})>0$ implies $\mathcal{N}'(r_{\mathrm{sp}})<0$ (stable static sphere), whereas $\mathcal{V}''_{\mathrm{eff}}(r_{\mathrm{sp}})<0$ implies $\mathcal{N}'(r_{\mathrm{sp}})>0$ (unstable static sphere), and $\mathcal{V}''_{\mathrm{eff}}(r_{\mathrm{sp}})=0$ implies $\mathcal{N}'(r_{\mathrm{sp}})=0$ (degenerate static sphere\footnote{A degenerate static sphere can arise when a stable and an unstable branch merge~\cite{Wei:2023bgp}. In such a merger, the opposite winding numbers, $+1$ and $-1$, sum to zero.}). 
	
	Equation (\ref{N=0}) therefore yields
	\begin{align}
		\label{p(rsp)}
		p(r_{\mathrm{sp}})=-\frac{m(r_{\mathrm{sp}})}{4\pi r_{\mathrm{sp}}^3}\;,
	\end{align}
	and, since by virtue of the WEC ($\rho\ge 0$) the mass $m(r_{\mathrm{sp}})=\frac{1}{2}r_H+\int_{r_H}^{r_{\mathrm{sp}}}4\pi r'^2\rho(r')dr'\ge r_H/2>0$, there must be a negative radial pressure (tension) at any static sphere.
	
	To examine the behavior of $\mathcal{N}(r)$ in the vicinity of a static sphere and to make contact with the energy conditions, we compute its derivative. After a straightforward but somewhat lengthy algebraic manipulation, we find
	\begin{align}
		\label{N1}
		\mathcal{N}'(r_{\mathrm{sp}})=-8\pi r_{\mathrm{sp}}(T+2\rho)=-8\pi r_{\mathrm{sp}}(\rho+p+2p_T)\;,
	\end{align}
	where we have used Eqs.~(\ref{mu1}), (\ref{p1}) and (\ref{N=0}). It is crucial to emphasize that Eq.~(\ref{N1}) is not an identity valid for all $r$; it holds specifically at the static sphere radius by virtue of the static sphere condition itself.
	
	The combination $\rho+p+2p_T$ is precisely the quantity appearing in the SEC. Hence, at a static sphere the sign of $\mathcal{N}'(r_\mathrm{sp})$ directly signals the sign of this combination:  assuming the WEC always holds, $\mathcal{N}'(r_{\mathrm{sp}})>0$ implies $\rho+p+2p_T<0$ (SEC violation), whereas $\mathcal{N}'(r_{\mathrm{sp}})\le 0$ implies $\rho+p+2p_T\ge 0$ (SEC holds).
	
	\begin{figure}[H]
		\centering
		\subfigure{
			\begin{minipage}[t]{0.5\linewidth}
				\centering
				\includegraphics[width=3.5in]{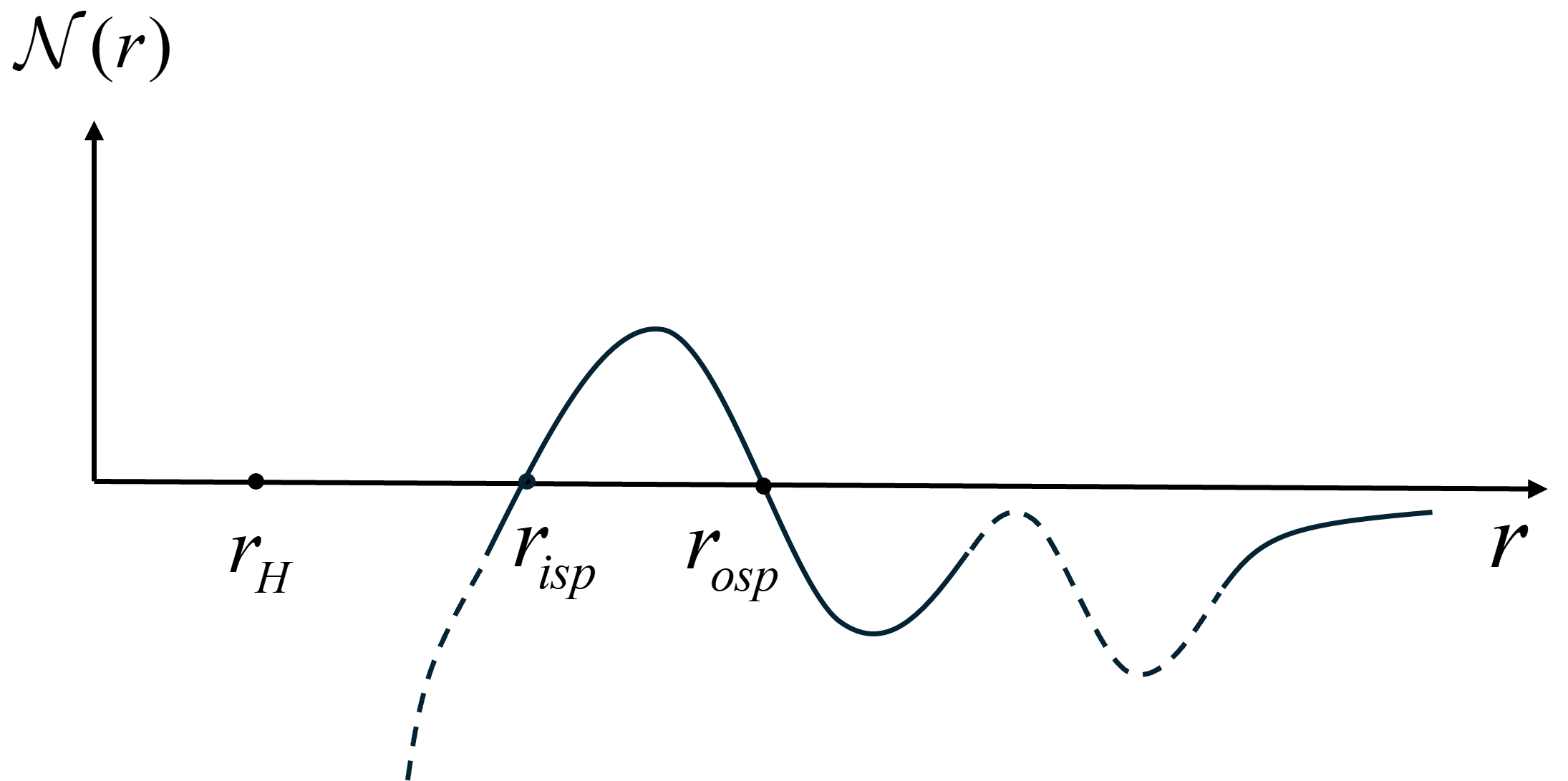}
				\begin{center}
					(a) $r_{\mathrm{isp}}$ denotes the inner (unstable) static sphere, $r_{\mathrm{osp}}$ denotes the outer (stable) static sphere.
				\end{center}
				\label{1a}
			\end{minipage}%
		}%
		\subfigure{
			\begin{minipage}[t]{0.5\linewidth}
				\centering
				\includegraphics[width=2.6in]{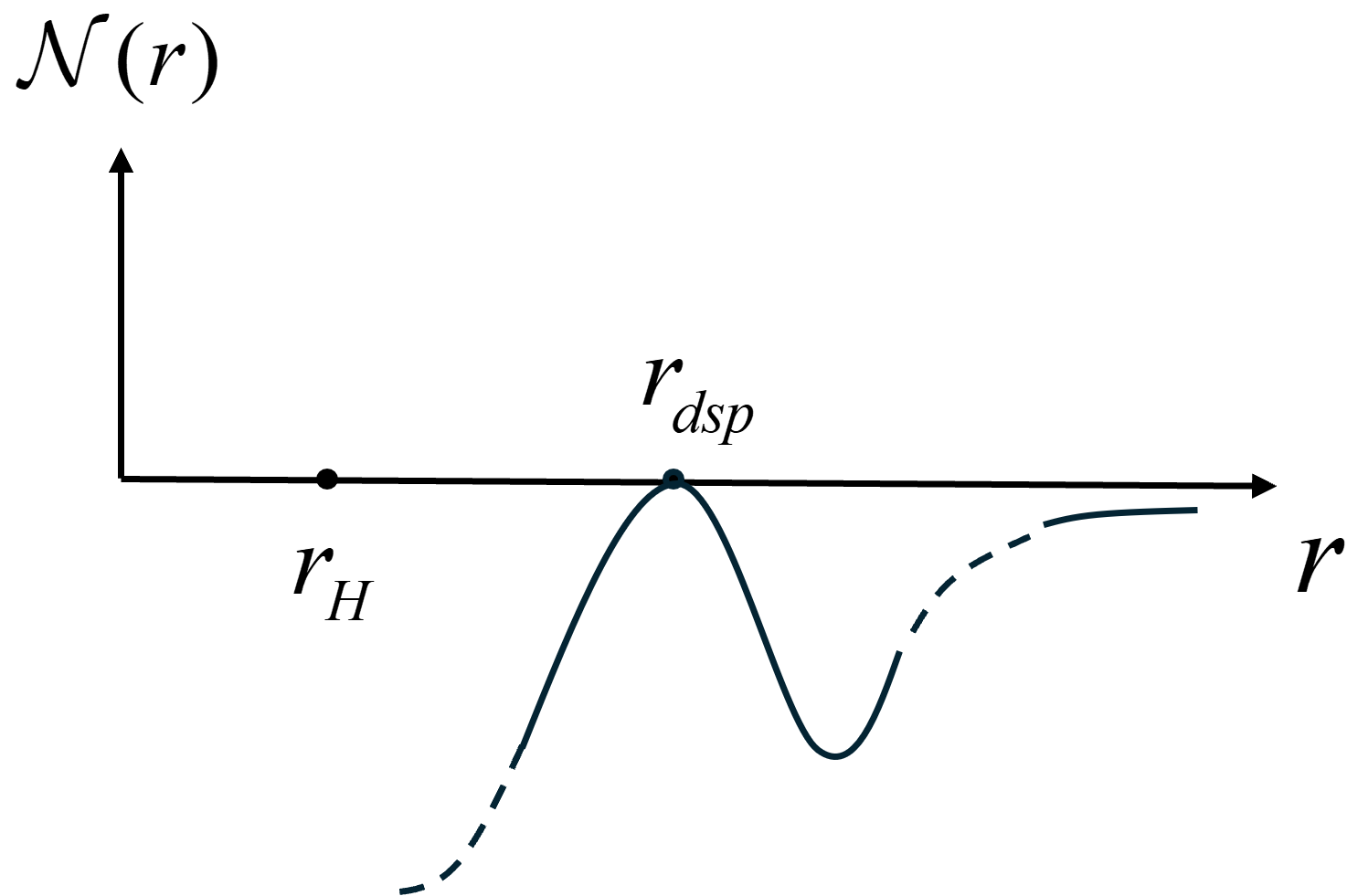}
				\begin{center}
					(b) $r_{\mathrm{dsp}}$ denotes a degenerate static sphere.
				\end{center}
				\label{1b}
			\end{minipage}%
		}%
		\centering
		\caption{Behavior of $\mathcal{N}(r)$ when static spheres exist. The figure is illustrative only; in general, multiple static sphere pairs, degenerate static spheres, or both may exist.}
		\label{1}
	\end{figure}
	
	We assume a non-extremal horizon, the WEC throughout the exterior, and a finite number of static spheres outside the horizon. We also assume that $\mathcal{N}(r)$ is continuously differentiable. We now establish the global behavior of $\mathcal{N}(r)$. At the horizon, using Eqs.~(\ref{murH}), (\ref{prH}) and (\ref{N}), we find
	\begin{align}
		\label{rh}
		\mathcal{N}(r_H)=-1-8\pi r^2_Hp(r_H)=8\pi r^2_H\rho(r_H)-1\leq 0\;.
	\end{align}
	Using Eq.~(\ref{N}) and the conditions (\ref{r3rho}) and (\ref{r3p}), one finds $\mathcal{N}(r)\sim -\frac{2M}{r}$ as $r\to\infty$, where $M$ is the ADM mass; hence
	\begin{align}
		\label{Rinfty}
		\mathcal{N}(r\rightarrow \infty)=0^-\;.
	\end{align}
	Thus, under these assumptions, $\mathcal{N}(r)$ is negative near the horizon and at sufficiently large radius. For a static sphere to exist (i.e., for $\mathcal{N}=0$ at some finite $r>r_H$), as shown in Fig.~\ref{1}, the function $\mathcal{N}(r)$ must attain a non-negative value somewhere between $r_H$ and infinity. This requires that $\mathcal{N}(r)$ increase from its non-positive horizon value to a non-negative maximum, implying that there exists an interval where $\mathcal{N}'(r)\ge 0$.
	
	To connect this with the energy conditions, we note that Eq.~(\ref{N1}) holds at any static sphere. For a pair of non-degenerate static spheres, the inner (unstable) one has $\mathcal{N}'>0$, implying $\rho+p+2p_T<0$ there; by continuity, this inequality must be valid in some open region around it, signaling a violation of the SEC. The outer (stable) sphere has $\mathcal{N}'<0$, so that $\rho+p+2p_T> 0$. At a degenerate static sphere, $\mathcal{N}=\mathcal{N}'=0$ and Eq.~(\ref{N1}) gives $\rho+p+2p_T=0$ (see Fig.~\ref{1}).
	
	Therefore, we arrive at one of our central conclusions: If all exterior static spheres are non-degenerate, i.e., $\mathcal{N}'(r_{\mathrm{sp}})\ne0$, the negativity of $\mathcal{N}$ near the horizon and at large radius implies that they occur in an even number, with alternating stability, since $\mathcal{N}$ changes sign at each sphere. In this case, the existence of static spheres requires a violation of the strong energy condition in some finite region outside the event horizon. Non-degenerate static spheres must appear in pairs---one unstable (inner) and one stable (outer). Denoting the inner unstable sphere by $r_{\mathrm{isp}}$ and the outer stable one by $r_{\mathrm{osp}}$, the inner sphere is characterized by $\mathcal{N}(r_{\mathrm{isp}})=0$, $\mathcal{N}'(r_{\mathrm{isp}})>0$ and thus $\rho+p+2p_T<0$ (SEC violation), while the outer sphere satisfies $\mathcal{N}(r_{\mathrm{osp}})=0$, $\mathcal{N}'(r_{\mathrm{osp}})<0$, where $\rho+p+2p_T>0$ (SEC restored). In the degenerate case, the static sphere $r_{\mathrm{dsp}}$ satisfies $\mathcal{N}(r_{\mathrm{dsp}})=\mathcal{N}'(r_{\mathrm{dsp}})=0$, and Eq.~(\ref{N1}) gives $\rho+p+2p_T=0$ at that sphere\footnote{It is worth emphasizing that the above argument does not, by itself, exclude the possibility that $\mathcal{N}(r)$ possesses more than two zeros, corresponding, in the non-degenerate case, to more than one pair of static spheres (e.g., four, six, eight, $\ldots$). When degenerate zeros are present, the number of distinct zeros may be odd. If all exterior static spheres are non-degenerate, successive zeros can be grouped into pairs of opposite stability: the innermost zero is unstable ($\mathcal{N}'>0$, SEC violated), and the outermost zero is stable ($\mathcal{N}'<0$, SEC restored).}.
	
	\section{Upper Bound on the Radius of the Innermost Static Sphere}\label{section4}
	
	We now focus on the innermost exterior static sphere, with $r^-_{\mathrm{sp}}>r_H$, for a non-extremal black hole. In the generic paired case, the inner static sphere is unstable, and we still have $\mathcal{N}'(r^-_{\mathrm{sp}})>0$ (since the first root is non-degenerate and $\mathcal{N}<0$ immediately to its left). From Eq.~(\ref{N1}) the SEC is violated at that point. To obtain a quantitative bound, we introduce the auxiliary function
	\begin{align}
		F(r)\equiv r\mathcal{N}(r)=-2m(r)-8\pi r^3 p(r)\;.
	\end{align}
	Using Eqs.~(\ref{m1}), (\ref{p1}) and (\ref{T}), we derive the exact linear differential equation
	\begin{align}
		\label{F1}
		F'(r)+\frac{4\pi r(\rho+p)}{\mu}F(r)=-8\pi r^2(\rho+p+2p_T)\;.
	\end{align}
	
	Consider the interval $[r_H,r^-_{\mathrm{sp}}]$ where $r^-_{\mathrm{sp}}$ is the innermost static sphere, i.e., the first exterior zero of $\mathcal{N}$ (and hence of $F$). We make the following physically motivated assumptions in this region:
	\begin{itemize}
		\item [(1)] Weak energy condition (WEC) holds. This guarantees $P(r)\equiv\frac{4\pi r(\rho+p)}{\mu}\ge 0$ because $\mu>0$ for $r>r_H$.
		\item [(2)] SEC is uniformly violated: there exists a constant $\kappa>0$ such that
		\begin{align}
			\label{eq:uniform}
			\rho+p+2p_T\le -\kappa,\quad \mathrm{for\;\; all}\quad r\in [r_H,r^-_{\mathrm{sp}}]\;.
		\end{align}
		Here $\kappa$ is a (positive) lower bound on $-(\rho+p+2p_T)$, i.e., a measure of the minimum strength of the SEC violation over this region. Consequently the source term in Eq.~(\ref{F1}) satisfies $Q(r)\equiv -8\pi r^2(\rho+p+2p_T)\ge 8\pi \kappa r^2$.
	\end{itemize}
	At $r=r^-_{\mathrm{sp}}$, Eq.~(\ref{F1}) and $F=0$ give $F'(r^-_{\mathrm{sp}})=Q(r^-_{\mathrm{sp}})\ge 8\pi\kappa(r^-_{\mathrm{sp}})^2>0$, confirming that the first root is non-degenerate and unstable under these assumptions. These conditions are mutually consistent. They describe, for example, matter that satisfies the stronger dominant energy condition ($|p|\le\rho\;, |p_T|\le \rho$,  which implies the WEC) while possessing sufficiently negative tangential pressures to break the SEC --- a situation commonly considered in models of dark energy, scalar-tensor gravity, or quantum-corrected black holes.
	
	Define $I(r)\equiv\int_{r_H}^{r}P(s)ds\ge 0$; by Eq.~(\ref{delta1}), $P=-\delta'$ and $I(r)=\delta(r_H)-\delta(r)$, which is finite at the horizon. Multiplying Eq.~(\ref{F1}) by $e^{I(r)}$ we obtain
	\begin{align}
		\frac{d}{dr}\bigg(F(r)e^{I(r)}\bigg)=Q(r)e^{I(r)}\;.
	\end{align}
	Integration from $r_H$ to $r^-_{\mathrm{sp}}$ yields
	\begin{align}
		\label{intF}
		F(r^-_{\mathrm{sp}})e^{I(r^-_{\mathrm{sp}})}-F(r_H)=\int_{r_H}^{r^-_{\mathrm{sp}}}Q(r)e^{I(r)}dr\;.
	\end{align}
	At the inner static sphere $F(r^-_{\mathrm{sp}})=0$. At the horizon, using Eqs.~(\ref{rhorH}), (\ref{prH}) and $m(r_H)=r_H/2$,
	\begin{align}
		F(r_H)=r_H\mathcal{N}(r_H)=r_H(8\pi r^2_H\rho(r_H)-1)\equiv -A\;,
	\end{align}
	where $A=r_H(1-8\pi r^2_H\rho(r_H))\ge 0$. Equation (\ref{intF}) therefore becomes
	\begin{align}
		\label{Aint}
		A=\int_{r_H}^{r^-_{\mathrm{sp}}}Q(r)e^{I(r)}dr\;.
	\end{align}
	Because $e^{I(r)}\ge 1$ and $Q(r)\ge 8\pi \kappa r^2$, we obtain the inequality
	\begin{align}
		\label{A}
		A\ge \int_{r_H}^{r^-_{\mathrm{sp}}}8\pi\kappa r^2dr=\frac{8\pi\kappa}{3}[(r^-_{\mathrm{sp}})^3-r_H^3]\;.
	\end{align}
	Solving for $r^-_{\mathrm{sp}}$ gives the upper bound\footnote{We emphasize that the bound (\ref{bound}) rests on the uniform-violation assumption (\ref{eq:uniform}), which supplies the pointwise lower bound $Q(r)\ge 8\pi\kappa r^2$ over the whole interval $[r_H,r^-_{\mathrm{sp}}]$. If the SEC violation is instead confined to a narrower region, this pointwise bound fails on part of the interval and the simple inequality (\ref{A}) no longer follows.}
	\begin{align}
		\label{bound}
		r^-_{\mathrm{sp}}\le \left[r_H^3+\frac{3r_H(1-8\pi r^2_H\rho(r_H))}{8\pi\kappa}\right]^{1/3}\;.
	\end{align}
	The bound (\ref{bound}) encodes a transparent physical picture:
	\begin{itemize}
		\item [(1)] The ``deficit'' $A$: At the horizon $\mathcal{N}(r_H)\le 0$. The quantity
		\begin{align}
			A=r_H(1-8\pi r_H^2\rho(r_H))
		\end{align}
		measures the deficit of $F=r\mathcal{N}$ at the horizon, where $F(r_H)=-A$. For a non-extremal black hole ($8\pi r_H^2\rho(r_H)<1$) we have $A>0$.
		\item [(2)] The ``driving force'' $\kappa$: This is the local strength of SEC violation. A larger $\kappa$ implies that the SEC-violating matter drives $\mathcal{N}$ upward more efficiently, resulting in a tighter upper bound and allowing the static sphere to lie closer to the horizon.
		\item [(3)] Volume interpretation: Rewriting Eq.~(\ref{A}) as
		\begin{align}
			\frac{4\pi}{3}\bigg[(r^-_{\mathrm{sp}})^3-r_H^3\bigg]\le\frac{A}{2\kappa}\;,
		\end{align}
		the left-hand side is the coordinate volume between the horizon and the inner static sphere. The bound states that this volume cannot exceed $A/(2\kappa)$. Thus, for a given deficit $A$ and SEC violation strength $\kappa$, the transition zone where $\mathcal{N}$ climbs from its horizon value to zero can be at most this thick. If the SEC-violating matter is weak ($\kappa$ small) the transition zone can be extended; if it is strong ($\kappa$ large) the inner static sphere must lie closer to the horizon.
		\item [(4)] Extremal case: For an extremal black hole, $A=0$, whereas Eq.~(\ref{Aint}) has a strictly positive right-hand side if $\kappa>0$ and $r^-_{\mathrm{sp}}>r_H$. Thus the stated assumptions exclude an exterior static sphere in this case. A horizon zero of $\mathcal{N}$ is not a timelike static orbit, since the static Killing vector is null at the horizon.
	\end{itemize}
	
	It is natural to ask whether an analogous upper bound can be derived for the outer (stable) static sphere, which would be dynamically relevant for accretion or thin-shell structures. Our method does not yield such a bound. The inner bound follows by anchoring $\mathcal{N}(r)$ at the horizon, where regularity forces $\mathcal{N}(r_H)\le 0$, and by treating the SEC violation as the source that drives $\mathcal{N}$ up to its first zero. The outer static sphere lacks an analogous horizon-deficit anchor: beyond it $\mathcal{N}(r)$ simply returns to $0^-$ at infinity, and its radius is governed by the detailed matter profile in the asymptotic region rather than by a universal deficit. Although the innermost static sphere is unstable, the bound on its radius remains physically meaningful. Its location is the first zero of $\mathcal{N}(r)$ and lies within the SEC-violating region, so the bound constrains how far from the horizon the first static sphere can appear and thus how far the SEC must be violated to bring $\mathcal{N}$ from its horizon value up to zero.
	
	\section{Discussion and Conclusion}\label{conclusion}
	
	In this work, we have conducted a local, analytic investigation of the conditions for the existence of static spheres in static, spherically symmetric, and asymptotically flat black hole spacetimes. By introducing the radial function $\mathcal{N}(r)$ constructed directly from the mass and radial pressure functions, we have proved that, under the WEC, any static sphere necessarily requires a negative radial pressure (i.e., tension). Under the assumptions stated in Sec.~\ref{section3}, including a finite number of exterior static spheres that are all non-degenerate, these spheres must occur in pairs---an inner unstable sphere and an outer stable one. This picture is fully consistent with the topological classification of static spheres introduced in Ref.~\cite{Wei:2023bgp}, where it was shown that for an asymptotically flat black hole the total topological number is zero, enforcing that static spheres appear in pairs with opposite stability indices. Our work provides a complementary, physically transparent picture: this pairing follows from the alternating sign changes of $\mathcal{N}(r)$ at the non-degenerate static spheres, together with its negativity near the non-extremal horizon and at large radius.
	
	We further established that, assuming the Weak Energy Condition (WEC) always holds, the inner (unstable) static sphere is universally characterized by a violation of the SEC, with $\rho+p+2p_T<0$, while at the outer (stable) static sphere the SEC is restored ($\rho+p+2p_T>0$). A degenerate static sphere satisfies $\rho+p+2p_T=0$, but its stability requires a higher-order analysis, and a pair-merger interpretation requires additional assumptions on a family of solutions. These results provide a direct, analytic link between the local energy-matter content and the global orbital structure of the spacetime.
	
	In addition, we derived a rigorous upper bound on the radius of the innermost static sphere, Eq.~(\ref{bound}), which is model-independent within general relativity, relying on the field equations, a regular non-extremal horizon, and the validity of the weak energy condition (WEC) and the uniform violation of the SEC in the region between the horizon and the static sphere. This generality makes the result applicable to a wide variety of black hole solutions of the Einstein field equations, including hairy black holes and those supported by SEC-violating matter fields (e.g., scalar fields, non-linear electrodynamics, or wave dark matter~\cite{Hui:2017ultra}), as long as the prescribed energy conditions are met. We stress that the derivation rests on the Einstein field equations (\ref{mu1})--(\ref{delta1}); the results therefore apply directly within general relativity. For modified theories of gravity, whose field equations differ from (\ref{mu1})--(\ref{delta1}), the present results do not carry over automatically, but may apply whenever the theory admits an equivalent description in terms of general relativity coupled to an effective energy-momentum tensor. The bound admits a transparent physical interpretation: the horizon deficit $A = r_H(1-8\pi r_H^2\rho(r_H))$ and the SEC violation strength $\kappa$ jointly limit the coordinate volume of the SEC-violating matter shell.
	
	These findings have several important implications. Our analysis reveals that any black hole spacetime that admits a pair of static spheres must host matter that violates the SEC near its (unstable) static sphere, which could be realized in suitable configurations of quantum fields, non-linear electrodynamics, or wave dark matter. Consequently, the observation or theoretical study of static spheres can be used to test and constrain non-standard compact objects and the matter content of the strong-field region. Moreover, measuring the location of the innermost static sphere, together with the horizon data and the stated assumptions, would place an upper limit on any admissible uniform lower bound $\kappa$ for the strength of SEC violation over the intervening region. Conversely, for a given matter model, the bound restricts the parameter space for which static spheres can exist outside the horizon.
	
	Several directions for future investigation are suggested by this work. First, the analysis should be extended to stationary and axisymmetric spacetimes, where the interplay between angular momentum and matter content could modify the existence and pairing of static rings or points, and could introduce richer degenerate structures. Second, it would be valuable to explore concrete asymptotically flat black hole solutions of general relativity with non-standard matter fields---such as certain scalar field configurations or non-linear electrodynamics, which violate the SEC while satisfying the WEC---that realize the paired static sphere scenario, and to examine how the degenerate static sphere manifests itself in such models, possibly through a bifurcation or phase transition. Finally, the relationship between static spheres and the violation of other energy conditions, such as the null or dominant energy conditions, could yield further insights into the allowable matter content around compact objects.
	
	\section*{Acknowledgement}
	This work is supported by the Research Start-up Funding of Chengdu University of Technology (Grant No. 10912-KYQD2022-09307).

\end{document}